\documentstyle[12pt]{article}

\def\be{\begin{equation}}
\def\ee{\end{equation}}
\def\bea{\begin{eqnarray}}
\def\eea{\end{eqnarray}}

\begin{document}

\begin{flushright}
hep-th/0007023
\end{flushright}

\pagestyle{plain}

\def\e{{\rm e}}
\def\haf{{\frac{1}{2}}}
\def\tr{{\rm Tr\;}}
\def\goes{\rightarrow}
\def\gym{g_{{}_{YM}}}
\def\hxmu{{\hat X^\mu}} \def\hxmd{{\hat X_\mu}}
\def\hxnu{{\hat X^\nu}} \def\hxnd{{\hat X_\nu}}
\def\nbnb{{N\times N}} \def\nbns{{n\times n}}
\def\dermu{{\partial^\mu}} \def\dermd{{\partial_\mu}}
\def\dernu{{\partial^\nu}} \def\dernd{{\partial_\nu}}
\def\omgd{{\Omega_{\mu\nu}}} \def\omgu{{\Omega^{\mu\nu}}}
\def\hqiu{{\hat q^i}}\def\hqid{{\hat q_i}}
\def\hqju{{\hat q^j}}\def\hqjd{{\hat q_j}}
\def\hpiu{{\hat p^i}}\def\hpid{{\hat p_i}}
\def\hpju{{\hat p^j}}\def\hpjd{{\hat p_j}}

\begin{center}
{\Large {\bf Gauge Symmetry As Symmetry Of \\
\vspace{.3cm} Matrix Coordinates}}

\vspace{.4cm}

Amir H. Fatollahi\footnote{fath@theory.ipm.ac.ir}

\vspace{.6 cm}

{\it Institute for Advanced Studies in Basic Sciences (IASBS),}\\
{\it P.O.Box 45195-159, Zanjan, Iran}\footnote{Permanent address.}

\vspace{.2cm}
{\it and}
\vspace{.2cm}

{\it Institute for Studies in Theoretical Physics and Mathematics (IPM),}\\
{\it P.O.Box 19395-5531, Tehran, Iran}

\vspace{.2cm}
{\it and}
\vspace{.2cm}

{\it High Energy Physics Division, Department of Physics,
University of Helsinki, and\\
Helsinki Institute of Physics, P.O.Box 9, FIN-00014
University of Helsinki, Finland}

\vskip .2 cm
\end{center}

\begin{abstract}
We propose a new point of view to gauge theories based on taking the
action of symmetry transformations directly on the coordinates of space.
Via this approach the gauge fields are not
introduced at the first step, and they can be interpreted as fluctuations
around some classical solutions of the model. The new point of view is
connected to the lattice formulation of gauge theories, and the parameter
of noncommutativity of coordinates appears as the lattice spacing
parameter. Through the statements concerning the continuum limit of
lattice gauge theories, this suggestion arises that the noncommutative
spaces are the natural ones to formulate gauge theories at strong
coupling. Via this point of view, a close relation between the large-$N$
limit of gauge theories and string theory can be manifested.
\end{abstract}

\vspace{.3cm}


\newpage

Recently a great attention is appeared in formulation and studying field
theories on non-commutative (NC) spaces \cite{9908142,072186,MSSW}. Apart
from abstract mathematical interests, the physical motivations in doing so
have been different. One of the original motivations has been to get
``finite" field theories via the intrinsic regularizations which are
encoded in some of NC spaces \cite{snyder,madore}. The other motivation
was coming from the unification aspects of theories on NC spaces. These
unification aspects has been the result of the ``algebraization" of
``space, geometry and their symmetries" via the approach of NC geometry
\cite{connes}. Interpreting Higgs field as a gauge field in discrete
direction of a two-sheet space \cite{connlott} and unifying gauge theories
with gravity \cite{conncham,wali} are examples of this view point to NC
spaces.

The other motivation comes back to the natural appearance of NC spaces in
some areas of physics, and the recent one in string theory. It has been
understood that string theory is involved by some kinds of
non-commutativities; two examples are, 1) the coordinates of bound-states
of $N$ D-branes are presented by $N\times N$ Hermitian matrices
\cite{9510135}, and 2) the longitudinal directions of D-branes in the
presence of B-field background appear to be NC, as are seen by the ends of
open strings \cite{CDS,jabbari,9908142}.

As mentioned in the above, one of the motivations to formulate theories on
NC spaces has been a unified treatment with the symmetries living in a
space and the space itself. One of the most important symmetries in
physical theories is gauge symmetry, and to be extreme in identifying the
space with its symmetries is to take the action of symmetry
transformations on the space. In usual gauge theories the action of the
symmetry transformations is defined on the gauge fields, $A^\mu$, but in
the new picture one takes the action on space, and to be more specific on
the ``coordinates" of space. It will be our main strategy in presenting a
new point of view to gauge theories. As it will be clear later, the main
tools and view points to different subjects and discussions here are
developed and coming form the D-branes of string theories \cite{Po2, Tay}.
Here we try to reorganize the facts and discussions to present a new
picture for gauge theories and see how the things should be by this
approach.  The action which we concern here is the Eguchi-Kawai one
\cite{egukaw}, but with a different interpretation on the configurations
which are described by the action. As we will see, the new interpretation
is sufficiently rich to recover some aspects of gauge theories which has
been already known as maybe some disjoint facts. It will be shown that the
new interpretation is related from one side to lattice formulation of
gauge theories \cite{willat}, and with a different representation is
connected to ordinary formulation of gauge theories. In relation with
lattice gauge theory the parameter of noncommutativity of coordinates
appears as the lattice spacing parameter. Through the statements
concerning the continuum limit of lattice gauge theories this suggestion
arises that the NC spaces are the natural ones to formulate gauge theories
at the strong coupling limit. Also the model can manifest a close relation
between the large-$N$ limit of gauge theories, known to be the theory of
``Feynman graphs" as the world-sheet of strings, and string theory
\cite{tooft}.

{\bf Note:} After completion of this work, I informed that lattice
regularization of NC gauge theory have been constructed as a natural
extension of Wilson's lattice gauge theory. Also the relation between
twisted Eguchi-Kawai model and a NC gauge theory have been studied 
\cite{amb}.


\vspace{.3cm}
{\large{\it The Model:}}

As mentioned in above, instead of introducing gauge fields, we define the
gauge symmetry transformations directly on the generators of displacement
in space, calling them ``coordinates" and representing by $\hxmu$
\cite{MSSW} \footnote{In \cite{MSSW} these object are called ``covariant
coordinates".}, and we assume to be $\nbnb$ Hermitian matrices. So to
describe the generators in an infinite volume these matrices should be
taken for $N\goes\infty$, even when they are used to formulate a finite
group gauge theory. So we take the definition of the gauge transformations
as:
\bea \label{symtra}
\hxmu\goes\hxmu{}'=\omega \hxmu \omega^\dagger,\;\;\;\;\mu=1,...,d,
\eea
where $\omega$ is an arbitrary unitary $\nbnb$ matrix
(so it belongs to a group, say $G$). This transformation
is the same of \cite{MSSW} but not in the infinitesimal
form. On the other hand, it is the same transformation
which acts on the coordinates of D-branes as $\nbnb$
Hermitian matrices (see e.g. \cite{fat021,fat241}). So
the coordinates in a space which contains
the bound-states of $N$ D-branes enjoy such a transformation.
Also if from the
first one chooses the matrices $\hxmu$s to be belong to
$L^2_\infty{(R\!\!\!\!R^d)}\otimes M_\nbns$
in the form $\hxmu=i\dermu\otimes{\bf 1}_n + \gym 1\otimes
A^\mu$ \cite{violet}, they will have the same behavior under gauge
transformations such as (\ref{symtra}). So we are not very
far from the usual language of gauge theories.

Our coordinates are matrices and NC, and as usual are accompanied
with a length scale which is the size that the NC effects appear.
Here we show this length scale by $\ell$. We define the unitary
matrices: \bea\label{udef} U_\mu\equiv\e^{i\ell \hxmu}, \eea
as the operators which acting on the states make the displacement $\ell$. With
the ideas coming from lattice gauge theory, and also reminding the
role of covariant derivatives as the tools of parallel
transformations, we define the objects: \bea\label{omgdef}
\omgd\equiv U_\mu U_\nu U_\mu^\dagger U_\nu^\dagger, \eea with
the property $\omgd^{-1}=\Omega_{\nu\mu}=\omgd^\dagger$. Then
the action of the model we take to be: \bea\label{action}
S=-\frac{1}{g^2}\sum_{\mu,\nu} \tr \omgd, \eea which via the $\tr$
is invariant under the transformation (\ref{symtra}). This action
is essentially the Eguchi-Kawai one \cite{egukaw}. In the context
of Eguchi-Kawai model the symmetry of the action is a global
symmetry, i.e. the symmetry transformations on the gauge fields
are space independent. But as we will see, interpreting $\hxmu$s
as space coordinates encodes sufficiently rich structure in
the model to extract gauge fields and their local symmetry
transformations as the same of usual formulation of gauge
theories. One may define something in analogy with the field
strength as: \bea \omgd\equiv\e^{- i\ell^2 F_{\mu\nu}}, \eea which
in small $\ell$ limit it takes the form: \bea
F^{\mu\nu}=-i[\hxmu,\hxnu]+ \haf\ell\bigg[\hxmu+\hxnu ,
[\hxmu,\hxnu]\bigg]+ O(\ell^2). \eea The action in small $\ell$
has the form as: \bea S|_{\ell\goes
0}=-\frac{1}{g^2}\sum_{\mu,\nu} \tr \bigg( 1-i\ell^2 F_{\mu\nu}
-\haf \ell^4 F_{\mu\nu}^2 +\;\cdots\bigg). \eea The linear term in
$F_{\mu\nu}$ does not have contribution to the action because it
is antisymmetric in $\mu\nu$ \footnote{It is not true that because
of $\tr$ the linear term can be ignored. For infinite dimensional
matrices one can get non-zero trace from a commutator.}. So for
small values of $\ell$ we have: \bea\label{smalll} S|_{\ell\goes
0}=-\frac{1}{2g^2}\ell^4\sum_{\mu,\nu} \tr [\hxmu,\hxnu]^2 +{\rm
const.\;term}+ O(\ell^5). \eea

The actions (\ref{action}) or (\ref{smalll}) are actions for the matrices describing
the space and its symmetries. Issues such as dynamical generation
of space and its dimension, and also the gauge group
via the Matrix Theory have been discussed in \cite{spadyn}\cite{9707162}.

\vspace{.3cm}
{\large{\it Relation To Lattice Gauge Theory (strong coupling):}}

The model described with action (\ref{action}) has already the
form of lattice gauge theory at large-$N$, called Eguchi-Kawai
model. Here we also want to mention the connection to lattice
gauge theory for finite groups. In fact the relation between NC
geometry and also NC differential geometry with lattice gauge theory
has already been established in the previous works
\cite{Balach,Dimakis}. Here we try to construct the relation
explicitly. Let us have a look to the action of lattice gauge
theory: \bea\label{lgt} S_{{\rm lgt}}=-\frac{1}{g^2}\sum_{\mu,\nu}
\sum_{\vec{i}} \tr \bigg(
\e^{iaA^\mu_{\vec{i}}}\;\e^{iaA^\nu_{\vec{i}+\mu}}\;\e^{-iaA^\mu_{\vec{i}+\nu}}\;
\e^{-iaA^\nu_{\vec{i}}}\bigg), \eea with $a$ as lattice spacing
parameter and $\vec i$ as a $d$-vector representing a site in the
$d$ dimensional lattice. Also we have used the symbol $\vec i +
\mu$ for $(i_1,\cdots,i_\mu+1,\cdots,i_d)$. To get a $U(m)$
lattice gauge theory, as the first step, take the Ansatz resulted
from $d$ times block-diagonalizations of the matrices $\hxmu$s,
with the size of the last block to be $m\times m$. So the action
takes the form: \bea S_{{\rm
blocked}}=-\frac{1}{g^2}\sum_{\mu,\nu} \sum_{\vec{i}} \tr \bigg(
\e^{i\ell\;\hat{x}^\mu_{\vec{i}}}\;\e^{i\ell\;\hat{x}^\nu_{\vec{i}}}
\;\e^{-i\ell\;\hat{x}^\mu_{\vec{i}}}\;
\e^{-i\ell\;\hat{x}^\nu_{\vec{i}}}\bigg), \eea which the index
$i_j$ in the vector $\vec i$ is counting the place of a block in
the $j$th step of block-diagonalizations. The $\tr$ above is for
the $U(m)$ structure of $\hat{x}^\mu_{\vec{i}}$ matrices. But this
action is still different from the lattice action (\ref{lgt}). To
make the exact correspondence we should do a slight modification
in one of the steps of block-diagonalizations. Firstly, take the
matrix $\Delta$ as: \bea
\Delta_{rs}&=&\delta_{r,s-1},\;\;\;\;\;\;\;\;\;\;\;\;\;\;\;\;\;\;\;\;\;\;\;
{\rm for\;infinite\;size},\nonumber\\
\Delta_{rs}&=&\delta_{r,s-1},\;\;\Delta_{p1}=1, \;\;\;\;\;\;\;{\rm
for\;size}\;p\times p, \eea with the properties
$\Delta^{-1}=\Delta^T=\Delta^\dagger$; so $\Delta$ is unitary. For
this matrix and a diagonal matrix $A$ we have: \bea \Delta\;{\rm
diag.}(a_1,\cdots,a_{p-1},a_p)\;\Delta^{-1}={\rm
diag.}(a_2,\cdots,a_p,a_1). \eea By using matrix $\Delta$ we
modify the block-diagonalizations mentioned above, by requesting
that in the $\mu$th step of diagonalizations of matrix $\hxmu$, it
picks up a $\Delta$ with appropriate size, as \bea \hxmu
\goes_{{}_{\!\!\!\!\!\!\!\!\!\!\!\!\!\!\!\mu{\rm th\;step}}}
\hat{x}^\mu\;\Delta. \eea So in two steps of $d$ steps two pairs
of $\Delta$ and $\Delta^{-1}$ appear around $\hxmu$ and $\hxnu$
matrices in the action, and this cause the appropriate shift in
the blocks to obtain the action of lattice gauge theory (\ref{lgt}).
In comparison with the lattice action one sees that the parameter
$\ell$ has appeared as the lattice spacing parameter. It means
that the lattice spacing parameter is a measure for appearing NC
effects \cite{Dimakis}. Based on the lattice calculations, one can
derive the relation between parameters $\ell$, the coupling
constant $g$ and the string tension $K$, and via this relation a
statement follows that the continuum limit of lattice gauge
theories are gained just at exactly zero coupling \cite{fat241,huang}. So
this suggestion arises that the strong coupling limit of gauge
theories will find a reasonable and natural formulation in NC spaces
(for more discussions on this point see \cite{fat241,fat021}). Also via this
explicit construction this observation is done that both the
structure of space (here a lattice) and also the gauge fields
living in the space can be extracted from the big matrices
$\hxmu$s. We see another example of this behavior in the relation
between the model and ordinary formulation of gauge theories.

\vspace{.3cm}
{\large{\it Relation To Ordinary Gauge Theory (weak coupling):}}

It is known that the classical action of lattice gauge theories at the
small lattice parameter is equivalent with the classical action of gauge
theories, so-called there, the weak coupling limit of lattice gauge theory
\cite{huang}. So up to know, by taking the limit $\ell\goes 0$ in the
action obtained in the previous section we can get the ordinary action of
gauge theories. In the following we give another presentation for this,
which of course it contains the procedure of going to continuum limit, but
a little implicitly. To get the ordinary gauge theory we use the
techniques which have been developed in constructing D-branes from Matrix
Theories \cite{BFSS,IKKT}. Here we just recall the construction and refer
the reader to literature (see e.g. \cite{BSS}). For large matrices one
always can find a set of matrix-pairs ($\hqiu$,$\hpiu$) with sizes
$n_i\times n_i$s so that: \bea [\hqiu,\hpju]=i\delta_{ij}{\bf 1}_{n_i}.
\eea The above commutator is not satisfied for finite dimensional
matrices. We assume the eigenvalues of $\hqiu$ and $\hpiu$ are distributed
uniformly in the interval $[\;0,\sqrt{2\pi n_i}\;]$. To get a $U(m)$ gauge
theory one can break the matrices $\hxmu$s with size $N$ to matrices with
sizes $n_i$s and $m$ such that: $N=m\cdot n_1n_2...n_{d/2}$ when $d$ is
even, and $N=m\cdot n_1n_2...n_{(d+1)/2}$ for $d$ odd, with the condition
$N,n_i\goes\infty$ and $m$ finite. On the other hand, it is easy to see
that matrices in the form: \bea \ell^2{\hat X}^{2i-1}_{{\rm
cl}}&=&\underbrace{{\bf 1}_{n_1}\otimes\cdots}_{i-1} \frac{\hqiu
L_i}{\sqrt{2\pi n_i}}\otimes\cdots\otimes{\bf 1}_{n_{d/2}}\otimes{\bf
1}_m,\nonumber\\ \ell^2{\hat X}^{2i}_{{\rm cl}}&=&\underbrace{{\bf
1}_{n_1}\otimes\cdots}_{i-1} \frac{\hpiu L_{i+1}}{\sqrt{2\pi
n_i}}\otimes\cdots\otimes{\bf 1}_{n_{d/2}}\otimes{\bf
1}_m,\;\;\;\;i=1,...,d/2, \eea for even $d$, and with an extra one as:
\bea \ell^2{\hat X}^d_{{\rm cl}}=\underbrace{{\bf
1}_{n_1}\otimes\cdots}_\frac{d-1}{2} \frac{{\hat q}^\frac{d+1}{2}
L_d}{\sqrt{2\pi n_\frac{d+1}{2}}}\otimes{\bf 1}_m, \eea for odd $d$, solve
the equations of motion derived from the action. Here $L_i$s have the
interpretation as the large radii of compactifications \cite{BFSS,IKKT}.
By the equations of motion for $n_i$s one obtains \cite{IKKT}: \bea
\frac{L_iL_{i+1}}{2\pi n_i}\sim \ell^2. \eea By admitting fluctuations
around classical solutions, one can write: \bea \hxmu=\hxmu_{{\rm cl}}+
\gym A^\mu, \eea with $A_\mu$s as $\nbnb$ Hermitian matrices and functions
of $(\hqiu,\hpiu)$ matrices, also with the same structure of matrices
$\hxmd$s. By inserting $\hxmd$s and expanding the action in the $\ell\goes
0$ limit up to second order of fluctuations, and with identifications
\cite{BSS,IKKT,BFSS}: \bea &~&[\hpid,*]\sim i\partial_{2i-1}*,\nonumber\\
&~& [\hqid,*]\sim i\partial_{2i}*,\nonumber\\ &~&\tr (\cdots)\goes \int
d^dx \;(\cdots), \eea one recovers the ordinary action for $U(m)$ gauge
theory. The coupling constant of the resulted gauge theory is found to be
$\gym^2\sim \ell^{d-4} g^2$, which in the limit of small $\ell$ and for
$d\geq 4$ the theory corresponds to the weak coupling limit.

\vspace{.3cm}
{\large{\it Large-$N$ Gauge Theory And String Theory:}}

It is known that in a diagrammatic representation, the partition function
of a gauge theory at large-$N$ is given by
$(\frac{1}{N})^{\rm genus}$ expansion,
with genus to be that of the ``big" Feynman graphs of the theory. Also it is
shown that the density of ``holes" (quark loops) in the graphs goes to zero with
$\frac{1}{N}$. So in the extreme large-$N$ limit the theory is
described by smooth graphs. By interpreting
$\frac{1}{N}$ as the coupling constant of a string theory, the
expansion mentioned above takes the form of the standard string
perturbation one \cite{tooft}. All of the
features mentioned here can be described by the point of view
proposed in this work. Firstly, at large-$N$ the action can take
the form of that of free strings.
We are thinking about smooth strings, so we take $\ell\goes 0$ and
$N\goes\infty$. So the action becomes:
\bea S|_{\ell\goes
0} = -\frac{1}{2\ell^4 g^2} \sum_{\mu,\nu}\tr [\hxmu,\hxnu]^2
-\frac{1}{g^2} \tr {\bf 1}_N, \eea
which we have applied the replacement $\hxmu\goes
\hxmu/\ell^2$ and so the new $\hxmu$ has the length dimension.
To get the free strings one use the map between the matrix
variables $(\hat q,\hat p)$ and continuous phase space variables
$(\sigma_1, \sigma_2)$ as \cite{IKKT,BFSS,zachos}:
\bea
&~&\tr(\cdots)\goes  \int d^2\sigma \sqrt{\det g_{rs}} \;(\cdots),\nonumber\\
&~&\;\;\left[A,B\right] \goes \{A,B\}_{{\rm PB}},\;\;\;\;\;\; \left[\hat q,\hat p\right]=i\goes
\{\sigma_1,\sigma_2\}_{{\rm PB}}=1/\sqrt{\det g_{rs}},
 \nonumber\\
&~&\;\;\;\;\left[\hat{p},*\right] \goes
i\partial_1*,\;\;\;\;\;\;\;\;\;\;\;\;\;\;
\;\;\;\;\;\;\left[\hat{q},*\right] \goes i\partial_2*, \eea with
the definition for Poisson bracket as $\{A, B\}_{{\rm
PB}}=\frac{1}{\sqrt{\det g_{rs}}} \epsilon^{rs}\partial_rA
\partial_rB$, ($r,s=1,2$). By these replacements one gets the
action of free strings in the Schild form \cite{schild,IKKT}. Also
by solving the equation of motion for $\sqrt{\det g_{rs}}$ and
inserting the solution in the action one can obtain the Nambu-Goto
action.

The issue of interaction is more subtle, and also has been
approached previously \cite{FKKT}. It is shown that the
$\frac{1}{N}$ expansion for this action corresponds to
perturbation theory of strings by reproducing the light-cone
string field theory through the Schwinger-Dyson equations.

\vspace{.3cm}
{\bf Acknowledgement:} This work was supported partially by the Academy of
Finland under the Project No. 163394. The author is grateful to M.M.
Sheikh-Jabbari for his comment on the first version.

\end{document}